\documentclass[runningheads]{llncs}
\usepackage[utf8]{inputenc} 
\usepackage[T1]{fontenc}    
\usepackage{hyperref}       
\usepackage{url}            
\usepackage[table]{xcolor}
\usepackage{amsfonts}       
\usepackage{nicefrac}       
\usepackage{microtype}      
\usepackage{orcidlink}
\usepackage{amssymb}
\usepackage{algorithm}
\usepackage{algpseudocode}
\usepackage{subcaption}
\usepackage{amsmath}
\usepackage{rotating}
\usepackage{makecell}
\usepackage{multirow}
\usepackage{enumitem}
\usepackage{tikz}
\usetikzlibrary{tikzmark}
\usepackage{booktabs} 
\usepackage{graphicx,verbatim}
\usepackage{arydshln}
\usepackage{tabularx} 
\usepackage{tabulary}
\usepackage{colortbl}
\usepackage{pifont}
\definecolor{cvprblue}{rgb}{0.21,0.49,0.74}
\definecolor{lightbrown}{rgb}{0.996,0.910,0.784}
\definecolor{darkbrown}{rgb}{0.996,0.816,0.647}

\renewcommand{\thefootnote}{\arabic{footnote}}

\begin{document}

\title{{Neural Finite-State Machines for Surgical Phase Recognition}}
\author{
Hao Ding$^{2,1,*}$ \and Zhongpai Gao$^{1, \dagger}$ \and Benjamin Planche$^1$ \and Tianyu Luan$^3$, Abhishek Sharma$^1$ \and Meng Zheng$^1$ \and Ange Lou$^4$ \and Terrence Chen$^1$ \and Mathias Unberath$^2$ \and Ziyan Wu$^1$}

\institute{$^1$United Imaging Intelligence, Boston MA, USA\\
$^2$Johns Hopkins University, Baltimore, MD, USA\\
$^3$State University of New York at Buffalo, Buffalo, NY, USA\\
$^4$Vanderbilt University, Nashville, TN, USA\\
\email{\{hding15,uberath\}@jhu.edu, \{first.last\}@uii-ai.com}}

%
%
%
%

\maketitle              
\begin{abstract}
Surgical phase recognition (SPR) is crucial for applications in workflow optimization, performance evaluation, and real-time intervention guidance. However, current deep learning models often struggle with fragmented predictions, failing to capture the sequential nature of surgical workflows. We propose the Neural Finite-State Machine (NFSM), a novel approach that enforces temporal coherence by integrating classical state-transition priors with modern neural networks.
NFSM leverages learnable global state embeddings as unique phase identifiers and dynamic transition tables to model phase-to-phase progressions. Additionally, a future phase forecasting mechanism employs repeated frame padding to anticipate upcoming transitions. Implemented as a plug-and-play module, NFSM can be integrated into existing SPR pipelines without changing their core architectures.
We demonstrate state-of-the-art performance across multiple benchmarks, including a significant improvement on the BernBypass70 dataset—raising video-level accuracy by 0.9 points and phase-level precision, recall, F1-score, and mAP by 3.8, 3.1, 3.3, and 4.1, respectively. Ablation studies confirm each component’s effectiveness and the module’s adaptability to various architectures. By unifying finite-state principles with deep learning, NFSM offers a robust path toward consistent, long-term surgical video analysis.

\keywords{Surgical Workflow Analysis  \and Surgical Video Analysis.}

\end{abstract}

\let\thefootnote\relax\footnotetext{\textsuperscript{*} This work was primarily carried out during the internship of Hao Ding at United Imaging Intelligence, Boston MA, USA.}
\let\thefootnote\relax\footnotetext{\textsuperscript{$\dagger$} Corresponding author.}


\section{Introduction}
\label{section:intro}

Phase recognition is fundamental to analyzing structured procedural videos, particularly in surgical operations. Accurate identification of procedure phases enables critical applications including surgical workflow optimization, performance assessment, and real-time intervention guidance. We focus on surgical phase recognition (SPR), an increasingly important challenge at the intersection of computer vision and surgical data science.

\begin{figure}[t]
    \centering
    \includegraphics[width=1.0\textwidth]{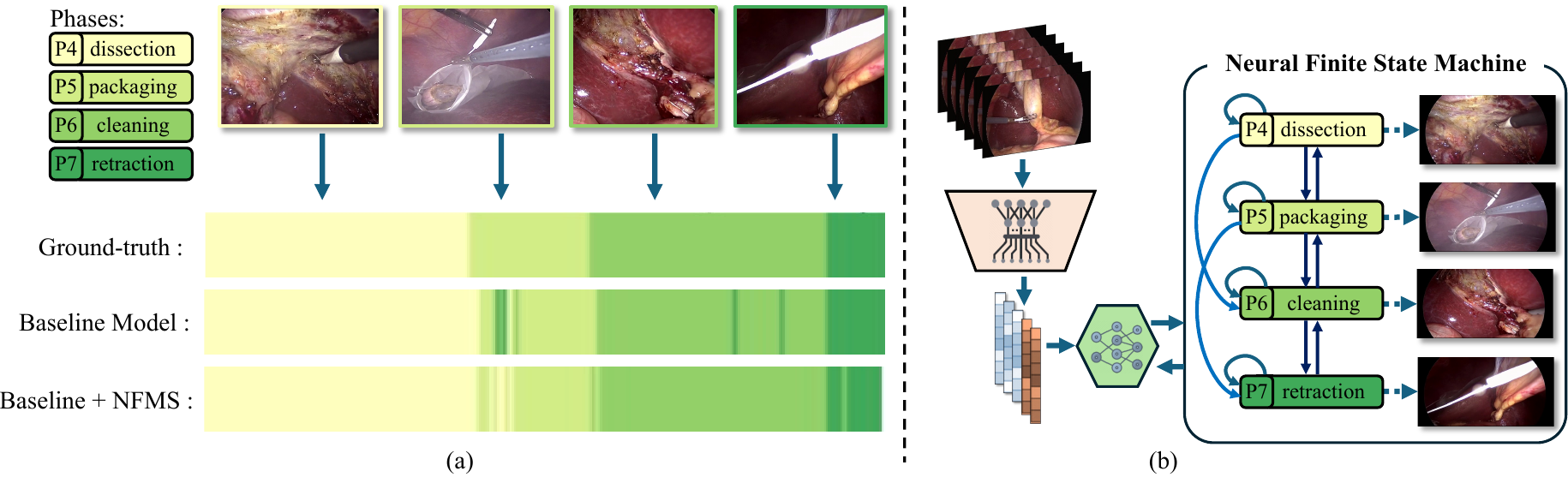}
    \caption{(a) Visualization of phase recognition consistency in surgical videos. While baseline methods misclassify isolated movements as phase transitions, our neural finite-state machine maintains phase coherence across longer durations. (b) Architecture overview showing the integration of our neural finite-state module.}
    \label{fig:finite_state_machine}
\end{figure}

The evolution of surgical phase recognition has been marked by increasingly sophisticated neural architectures. Initial attempts~\cite{TwinandaSMMMP17endonet,JinLDCQFH20mrcnet,JinD0YQFH18sv-rcnet,YiJ19hfom,GaoJDH20treesearch,JinLCZDH21tmrnet,czempiel2020tecno} applies CNN-based feature extraction with various temporal modeling approaches like long short term memory (LSTM)~\cite{hochreiter1997long} and temporal convolutional network (TCN) ~\cite{LeaVRH16tcn}. The advent of transformers~\cite{VaswaniSPUJGKP17transfomer} and ViT~\cite{DosovitskiyB0WZ21vit} enabled parallel processing of longer sequences. This led to stronger architectures~\cite{GaoJLDH21transsvnet,Liu23Lovit,valderrama2020tapir,GirdharG21avt,KilleenZMATOU23_Pelphix,yang2024surgformer,ding2024towards} with capacity to extract features directly from complex surgical videos clips.
However, modeling long-term dependencies remains a fundamental challenge in this learning-based video understanding. This limitation becomes particularly acute in surgical videos, which often span hours, making it difficult for current approaches to maintain consistent phase recognition. Figure~\ref{fig:finite_state_machine}\textcolor{black}{a} illustrates this challenge, where baseline methods (second bar) exhibit fragmented predictions within the surgical phase of packaging (P5), misclassifying temporary movements as phase transitions. While this reflects the difficulty in maintaining temporal consistency, we observe that surgical procedures follow well-defined workflows with predictable phase transitions. This property is well recognized by classical HMM-based approaches~\cite{BlumPFN08HMM,BhatiaOXH07RL_video_hmm,CadeneRTC16CNN_HMM_SMOOTHING,PadoyBFBN08_online,BlumFN10,PadoyBAFBN12} that modeled surgical procedures as finite-state machines, defining transitions between predefined states and their observations. Although their performances are not comparable to the deep learning-based method due to lack of strong visual feature extractor, modeling state transition priors can still be informative in capturing essential long-term procedural understanding.

Thus, we introduce a neural finite-state machine (NFSM) to enhance long-term consistency in surgical phase recognition (SPR). Our approach augments existing methods through systematic procedural understanding via several key components. First, we utilize learnable global embeddings to distinguish between surgical phases, essentially creating unique phase identifiers that capture specific characteristics of each phase. Second, through an attention-based mechanism, these embeddings generate a dynamic transition table that predicts phase-to-phase transitions as surgery progresses. Third, this dynamic table actively interacts with keyframe predictions, allowing the model to adapt its focus based on phase-specific information during training and inference.
The NFSM module can be used as a plug-and-play module that seamlessly integrates with state-of-the-art models through an addition attention block for future forecasting. We innovatively use repeated current frame padding as pseudo embeddings for future frames, enabling the model to learn and anticipate phase transitions in ongoing surgeries. During inference, the model combines predicted probabilities with transition probabilities computed from the dynamic transition tables to enhance prediction accuracy.

Our evaluation on surgical benchmarks like BernBypass70~\cite{Lavanchy2024}, Cholec80~\cite{TwinandaSMMMP17endonet}, and AutoLaparo~\cite{wang2022autolaparo} demonstrates state-of-the-art performance across comprehensive metrics. Especially on the most challenging BernBypass70 benchmark with largest scale among all three benchmark, NFSM improves baseline Surgformer model 0.9 video-level accuracy and 3.8, 3.1, 3.3, and 4.1 phase-level precision, recall, F1-scores, and mAP, respectively. Comprehensive ablation study shows the effectiveness of all proposed components as well as its generalizability to different architectures and adaptability to scenarios where finetuning is not applicable.


This work’s primary contributions can be summarized as follows:

\begin{itemize}[topsep=0pt]
  \item We introduce a novel Neural Finite-State Machine (NFSM) that bridges the gap between classical finite-state modeling and modern deep learning architectures, thereby enabling more temporally consistent SPR.
  \item NFSM leverages learnable global state embeddings and dynamic transition tables to capture structured procedural knowledge, enhancing both temporal coherence and predictive accuracy.
  \item As a plug-and-play module, NFSM integrates seamlessly into existing methods with transition-aware training and inference, requiring minimal modifications while significantly improving performance.
  \item We demonstrate state-of-the-art results on multiple benchmarks, including the challenging BernBypass70 dataset, and present extensive ablation studies that validate NFSM’s adaptability and generalizability across diverse architectures and training regimes.
\end{itemize}
    
\section{Neural Finite-State Machine (NFSM)}
\label{section:method}

In this section, we present our approach in three parts. The first part introduce the overall framework that integrated NFSM as a plug-and-play module. The second and third parts introduce transition-aware training and inference.

\begin{figure}[t]
  \centering
   \includegraphics[width=1.0\textwidth]{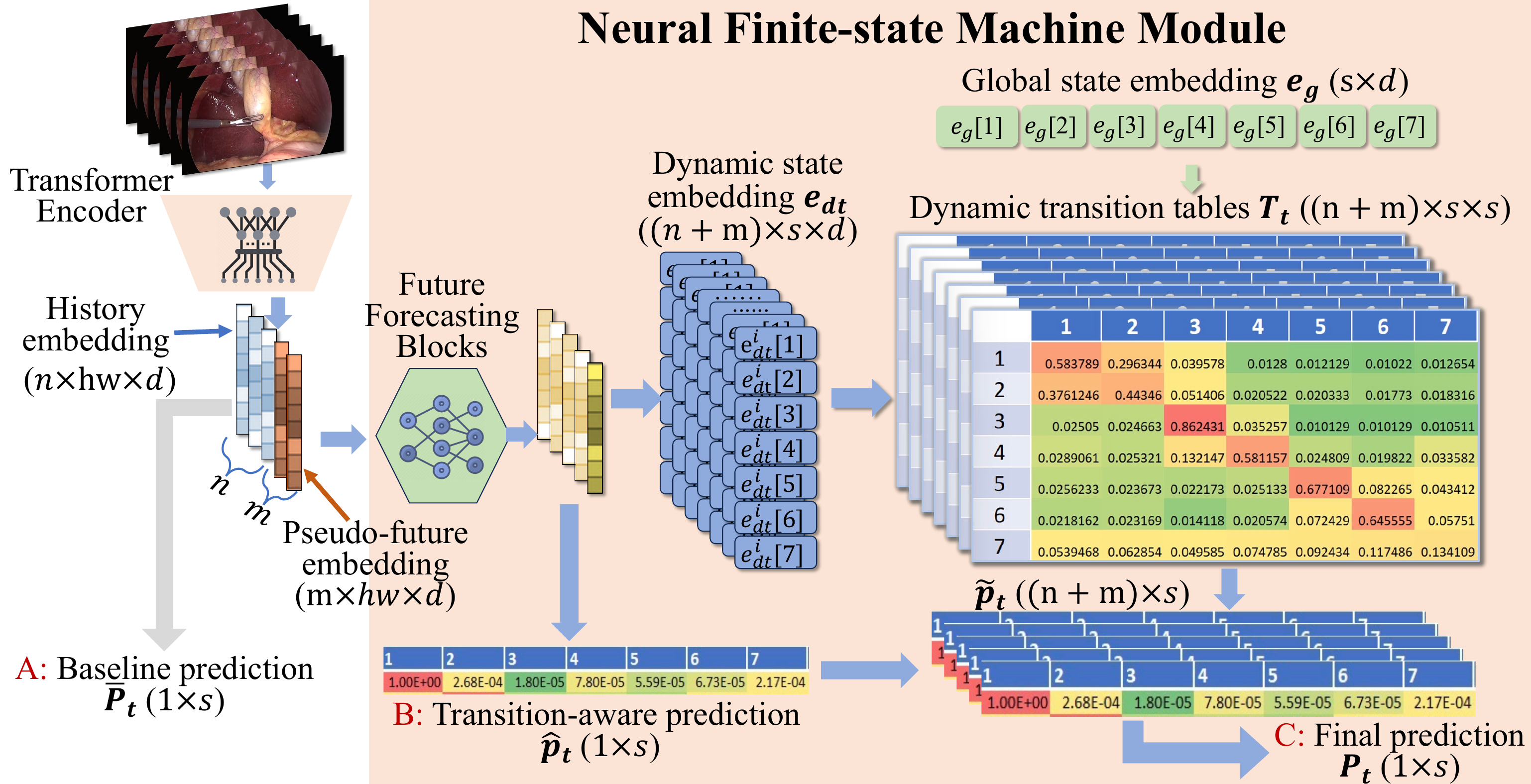}
   \caption{Overview of the Architecture.}
   \label{fig:architecture}
\end{figure}

\paragraph{\bf{Overall Framework.}} As shown in Figure~\ref{fig:architecture}, the overall framework takes a sequence of video frames $\{t-n+1, t-n+2, \ldots, t\}$ as input, where $n$ represents the window size. Using a transformer backbone, it extracts spatio-temporal features and produces history embedding ($(n \times hw \times d)$-dimensional). We duplicate the current frame's embedding $m$ times ($(m\times hw \times d)$-dimensional). This pseudo-future embedding enables dynamic transition table generation for future state prediction in online applications. Note that we can simply use real future embeddings for offline applications. The combined embedding is then fed into the future forecasting blocks made of two attention blocks. The future forecasting blocks output features after spatial pooling and then have two predictions: current frame state probabilities $\hat{p}_t\in\mathbb{R}^{1\times s}$ through additional temporal pooling and dynamic state embeddings $e_{dt}\in\mathbb{R}^{(n+m)\times s\cdot d}$ reshaped to $e_{dt}\in\mathbb{R}^{(n+m)\times s\times d}$, where $s$ denotes the total number of states (phases). 

The NFSM module processes these dynamic state embeddings using a set of learnable global state embeddings $e_g\in\mathbb{R}^{s \times d}$. For each frame $i$ in the input sequence $\{t-n+1, t-n+2,\dots,t+m\}$, the module leverages the dynamic state embedding $e_{dt}^i$ and the global state embedding $e_g$ to compute dynamic transition tables $T_t^i$ via the following relation:
\begin{equation}
  T_t^i = \text{softmax}(\frac{e_{dt}^i \cdot e_g}{\sqrt{d}}),
  \label{eq:dynamic_transition_table}
\end{equation}
where the softmax function is applied to normalize the calculated values, ensuring they represent valid transition probabilities. The dynamic transition tables facilitate the computation of transition state probabilities for both historical and future frames relative to the current frame. This computation is performed by multiplying these transition tables $T_t^i$ with the phase prediction probabilities $\hat{p}_t$ of the current frame, yielding a comprehensive set of transition state probabilities $\tilde{p_t} = \{\tilde{p}_t^{t-n+1}, \tilde{p}_t^{t-n+2}, \dots, \tilde{p}_t^{t+m}\}$.

\paragraph{\bf{Transition-aware Training.}}
In the training phase, we apply dual supervisions: one on the direct prediction of state probabilities $\hat{p}_t$ and another on the transition state probabilities $\tilde{p_t}$ generated by the NFSM.  First, the loss $\mathcal{L}_c$ for the prediction state probabilities of the current frame is computed using the Kullback-Leibler (KL) divergence between the prediction state probabilities $\hat{p}_t$ and the one-hot encoded ground truth label $y_t$, formalized as
\begin{equation}
    \mathcal{L}_c = \sum_{j = 1}^{s} -y_t[j]\log(\hat{p}_t[j]).
\end{equation}
Second, we leverage history and future frame labels $\{f_{t-n+1}, f_{t-n+2}, \dots, f_{t+m}\}$ to serve as guidance to learn predictive capabilities based on the history and current frames. The loss $\mathcal{L}_{trans}$ for the transition state probabilities is defined as \begin{equation}
\mathcal{L}_{trans} = \frac{1}{n+m}\sum_{i=t-n+1}^{t+m}\sum_{j = 1}^{s} -y_i[j]\log(\tilde{p}_t^i[j]).
\end{equation}

The final loss $\mathcal{L}$ is then calculated as a weighted sum of these components, $\mathcal{L} = \mathcal{L}_t + \alpha \mathcal{L}_{trans}$, where $\alpha$ is a weighting factor balancing the contributions of direct prediction and transition state predictions. This strategy not only guides the NFSM to capture accurate phase-level representations by optimizing the alignment between the dynamic state and global state embeddings based on historical frames but also fosters the model's ability to project these embeddings into future frames, ensuring a comprehensive understanding and anticipation of surgical phase transitions.

\paragraph{\bf{Transition-aware Inference.}} Leveraging the learned transition dynamics, we introduce a \textit{transition-aware} inference strategy. During inference at given time $t$, the model processes a sequence of frames $\{f_{t-n+1}, f_{t-n+2}, \dots, f_t\}$, forecasting not only the current frame's prediction state probabilities $\hat{p}_t$ but also the future $m$ frames' transition state probabilities $\{\tilde{p}_t^{t+1}, \tilde{p}_t^{t+2}, \dots, \tilde{p}_t^{t+m}\}$ through the NFSM. This approach stores the transitioning state probabilities from the current state to future states, which also means we can aggregate the transition state probabilities from past $m$ frames to the current frame (note that the current frame is the future of the past $m$ frames) into a consolidated transition state probability $\tilde{p}^t$ for the current frame. Specifically, $\tilde{p}^t$ is calculated as the average of transition state probabilities from the past $m$ frames: $\tilde{p}^t = \frac{1}{m}\sum_{j = 1}^m \tilde{p}_{t-j}^t$. The final prediction $p_t$ for frame $t$ merges these insights, normalizing the element-wise multiplication of the direct transition-aware prediction $\hat{p}_t$ and the aggregated transition-state probabilities $\tilde{p}^t$, as $p_t=\text{normalized}(\hat{p}_t \odot \tilde{p}^t)$. In this way, the model exploits all relevant transition probabilities for the frame $t$, enriching the final prediction with a comprehensive temporal context.
\begin{table}[t]
  \centering
    \caption{Online surgical phase recognition results. The metrics in the original Surgformer paper~\cite{yang2024surgformer} differ due to a different unrelaxed calculation. Here, we adopt the metric definitions and implementations from SKiT~\cite{liu2023skit} for a fair comparison across methods.}
\resizebox{0.95\linewidth}{!}{
     \begin{tabular}{cllllll}
     \toprule
 \multirow{2}{*}{Datasets}  & \multirow{2}{*}{Methods}  & \multirow{2}{*}{\makecell{Video-level\\ Accuracy}} & \multicolumn{4}{c}{Phase-level} \\
 \cmidrule(lr){4-7}&   &  & Precision & Recall & Jaccard & mAP \\
 \midrule
& TeCNO ~\cite{czempiel2020tecno} 
& $83.8 \pm 13.6$ & $61.28$ & $62.81$ & $59.22$ & - \\
& MTMS-TCN ~\cite{Lavanchy2024} 
& $85.3 \pm 13.2$ & $64.6$ & $67.4$ & $62.4$ & - \\
\cdashline{2-7}
BernBypass70~\cite{Lavanchy2024}
& Surgformer~\cite{yang2024surgformer}
& $85.9 \pm 12.3$ & $66.1$ & $65.1$ & $62.5$ & 67.7 \\
& ~\cite{yang2024surgformer}  + NFSM (\textbf{Ours})
& \cellcolor{darkbrown}\bf{86.8 $\pm$ 12.3} 
&\cellcolor{darkbrown}\bf{69.9}
&\cellcolor{darkbrown}\bf{68.2}
&\cellcolor{darkbrown}\bf{65.8}
&\cellcolor{darkbrown}\bf{71.9}\\
\midrule
& TeSTra~\cite{zhao2022real}
& $90.1 \pm 7.6$ & $82.8$ & $83.8$ & $71.6$ & - \\
& Trans-SVNet~\cite{GaoJLDH21transsvnet}
& $89.1 \pm 6.6$ & $84.7$ & $83.6$ & $72.5$ & - \\
& LoViT~\cite{Liu23Lovit}  
& $91.5 \pm 6.1$ & $83.1$ & $86.5$ & $74.2$ & - \\
Cholec80~\cite{TwinandaSMMMP17endonet}& 
SKiT~\cite{liu2023skit}  
& \cellcolor{darkbrown}\bf{92.5 $\pm$ 5.1} & $84.6$ & \cellcolor{darkbrown}\bf{88.5} & $76.7$ & - \\
\cdashline{2-7}
&  Surgformer~\cite{yang2024surgformer} 
& $92.3 \pm 6.2$ & \cellcolor{lightbrown}87.1  & $87.6$ & \cellcolor{lightbrown}$77.8$ & 90.7 \\
& ~\cite{yang2024surgformer}  + NFSM (\textbf{Ours})
& \cellcolor{lightbrown}$92.3 \pm 5.4$  
& \cellcolor{darkbrown}\bf{87.3} 
& \cellcolor{lightbrown}$87.7$
& \cellcolor{darkbrown}\bf{78.0}
& \cellcolor{darkbrown}\bf{92.4} \\
\midrule
& TMRNet ~\cite{JinLCZDH21tmrnet} 
& $78.2$ & $66.0$ & $61.5$ & $49.6$ & - \\
& TeCNO ~\cite{czempiel2020tecno} 
& $77.3$ & $66.9$ & $64.6$ & $50.7$ & - \\
& Trans-SVNet~\cite{GaoJLDH21transsvnet}
& $78.3$ & $68.0$ & $62.2$ & $50.7$ & - \\
AutoLaparo~\cite{wang2022autolaparo}
& LoViT~\cite{Liu23Lovit}  
& $81.4 \pm 7.6$ & \cellcolor{darkbrown}\bf{85.1} & $65.9$ & $55.9$ & - \\
& SKiT~\cite{liu2023skit}  
& $82.9 \pm 6.8$ & 81.8 & $70.1$ & $59.9$ & - \\
\cdashline{2-7}
& Surgformer~\cite{yang2024surgformer}  
& \cellcolor{lightbrown}$86.1 \pm 7.3$ & 81.5 & \cellcolor{lightbrown}70.8 & \cellcolor{lightbrown}62.4 & 79.6 \\
& ~\cite{yang2024surgformer}  + NFSM (\textbf{Ours})
& \cellcolor{darkbrown}\bf{86.6 $\pm$ 6.2 } & \cellcolor{lightbrown}85.0 
& \cellcolor{darkbrown}\bf{71.5} 
& \cellcolor{darkbrown}\bf{63.7 }
& \cellcolor{darkbrown}\bf{82.7}\\
\bottomrule
\end{tabular}}
    
  \label{tab:benchmark}
\end{table}

\section{Experiments}
\label{section:experiment}

\subsection{Experimental Settings}


\subsubsection*{Datsets:} We evaluate our method on three datasets: 
(1) \textit{BernBypass70}~\cite{Lavanchy2024} contains 70 videos averaging 72 minutes. 
(2) \textit{Cholec80}~\cite{TwinandaSMMMP17endonet} includes 80 cholecystectomy videos divided into 7 phases, with a mean duration of 39min.
(3)  \textit{AutoLaparo}~\cite{wang2022autolaparo} comprises 21 hysterectomy videos, averaging 66 minutes and divided into 7 phases. 
All surgical videos are captured at $25$ frames per second (FPS) and subsequently downsampled to $1$ FPS to enable surgeons to accurately annotate specific surgical phases. To maintain experimental consistency, we strictly follow the data splits established in previous studies~\cite{TwinandaSMMMP17endonet,JinD0YQFH18sv-rcnet,JinLCZDH21tmrnet,GaoJLDH21transsvnet,valderrama2020tapir,liu2023skit,Lavanchy2024}. 


\subsubsection*{Evaluation Metrics.} Following the current practice of Cholec80 and AutoLaparo We employ four distinct metrics for surgical phase recognition evaluation: video-level \textit{accuracy}, phase-level \textit{precision}, \textit{recall}, and \textit{Jaccard}.
As all previous relaxed metric evaluations are based on the evaluation code~\cite{funke2023metrics} with issues, we only evaluate with non-relaxed metrics. Following SKiT~\cite{liu2023skit}, we first concatenate the predictions and ground truth labels from all videos into a single continuous sequence. We then compute the average performance per phase.
For the BernBypass70 dataset, we follow their evaluation metric~\cite{Lavanchy2024} with officially released code which consists of video-level \textit{accuracy} and phase-level \textit{precision}, \textit{recall}, and \textit{F1 score}. Unlike Cholec80 and AutoLaparo, the phase-level performance metrics are averaged across phases per video and then across videos.
We also include the phase-level mean average precision (mAP) applied in~\cite{valderrama2020tapir} with confidence-based sorting to capture more holistic evaluation for all benchmarks.
%

\begin{table*}[t]
  \centering
  \caption{Effectiveness of the proposed components for surgical phase recognition. A: baseline prediction $\bar{p}_t$; B: transition-aware prediction $\hat{p}_t$; C: full prediction $p_t$; \ding{55}: freeze baseline model (i.e., only train NSFM module). 
  }
\resizebox{1.0\linewidth}{!}{
     \begin{tabular}{ll|l|llll|l|llll}
     \toprule
      &  
      & \multicolumn{5}{c|}{AutoLaparo~\cite{wang2022autolaparo}} & \multicolumn{5}{c}{BernBypass70~\cite{Lavanchy2024}} \\
      \cmidrule(lr){3-12}
    & 
     & {Video-level} & \multicolumn{4}{c|}{Phase-level} 
     & {Video-level} & \multicolumn{4}{c}{Phase-level} \\
    &
     & \multicolumn{1}{c|}{Accuracy} & \multicolumn{1}{c}{Precision} & \multicolumn{1}{c}{Recall} & \multicolumn{1}{c}{Jaccard}& \multicolumn{1}{c|}{mAP}
     & \multicolumn{1}{c|}{Accuracy} & \multicolumn{1}{c}{Precision} & \multicolumn{1}{c}{Recall} & \multicolumn{1}{c}{Jaccard} & \multicolumn{1}{c}{mAP} \\
\midrule
& A
& $86.1$ & $81.5$ & $70.8$ & $62.4$ & 79.6
& $85.9$ & $66.1$ & $65.1$ & $62.5$ & 67.7
\\
\midrule
& B  
& 86.4 \textsubscript{(+0.3)}
& 83.7 \textsubscript{(+2.2)} 
& 71.5 \textsubscript{(+0.7)}
& 63.7 \textsubscript{(+1.3)}
& 81.9 \textsubscript{(+2.3)}

&86.7 \textsubscript{(+0.8)}
&69.3 \textsubscript{(+3.2)}
&68.4 \textsubscript{(+3.3)}
&65.8 \textsubscript{(+3.3)}
&71.1 \textsubscript{(+3.4)}\\
& C
& 86.6 \textsubscript{(+0.5)}
& 85.0 \textsubscript{(+3.5)} 
& 71.5 \textsubscript{(+0.7)}
& 63.7 \textsubscript{(+1.3)}
& 82.7 \textsubscript{(+3.1)}

&86.8 \textsubscript{(+0.9)}
&69.9 \textsubscript{(+3.8)}
&68.2 \textsubscript{(+3.1)}
&65.8 \textsubscript{(+3.3)}
&71.9 \textsubscript{(+4.2)}
\\
\midrule
 \ding{55} & B
& 86.1
& 81.5
& 72.3 \textsubscript{(+1.5)}
& 64.1 \textsubscript{(+1.7)}
& 82.7 \textsubscript{(+3.1)}

& 86.2 \textsubscript{(+0.3)}
& 68.3 \textsubscript{(+2.2)}
& 66.6 \textsubscript{(+1.7)}
& 64.0 \textsubscript{(+1.5)}
& 68.4 \textsubscript{(+0.7)}
\\
 \ding{55} & C
& 86.4 \textsubscript{(+0.3)}
& 84.5 \textsubscript{(+3.0)}
& 72.3 \textsubscript{(+1.4)}
& 64.1 \textsubscript{(+1.7)}
& 83.6 \textsubscript{(+4.0)}

& 86.3 \textsubscript{(+0.4)}
& 70.1 \textsubscript{(+4.0)}
& 68.3 \textsubscript{(+2.5)}
& 65.7 \textsubscript{(+3.2)}
& 69.4 \textsubscript{(+1.7)}
\\
\bottomrule
\end{tabular}}
  \label{tab:components}
\end{table*}

\begin{figure}[t]
    \centering
    \includegraphics[width=0.33\linewidth]{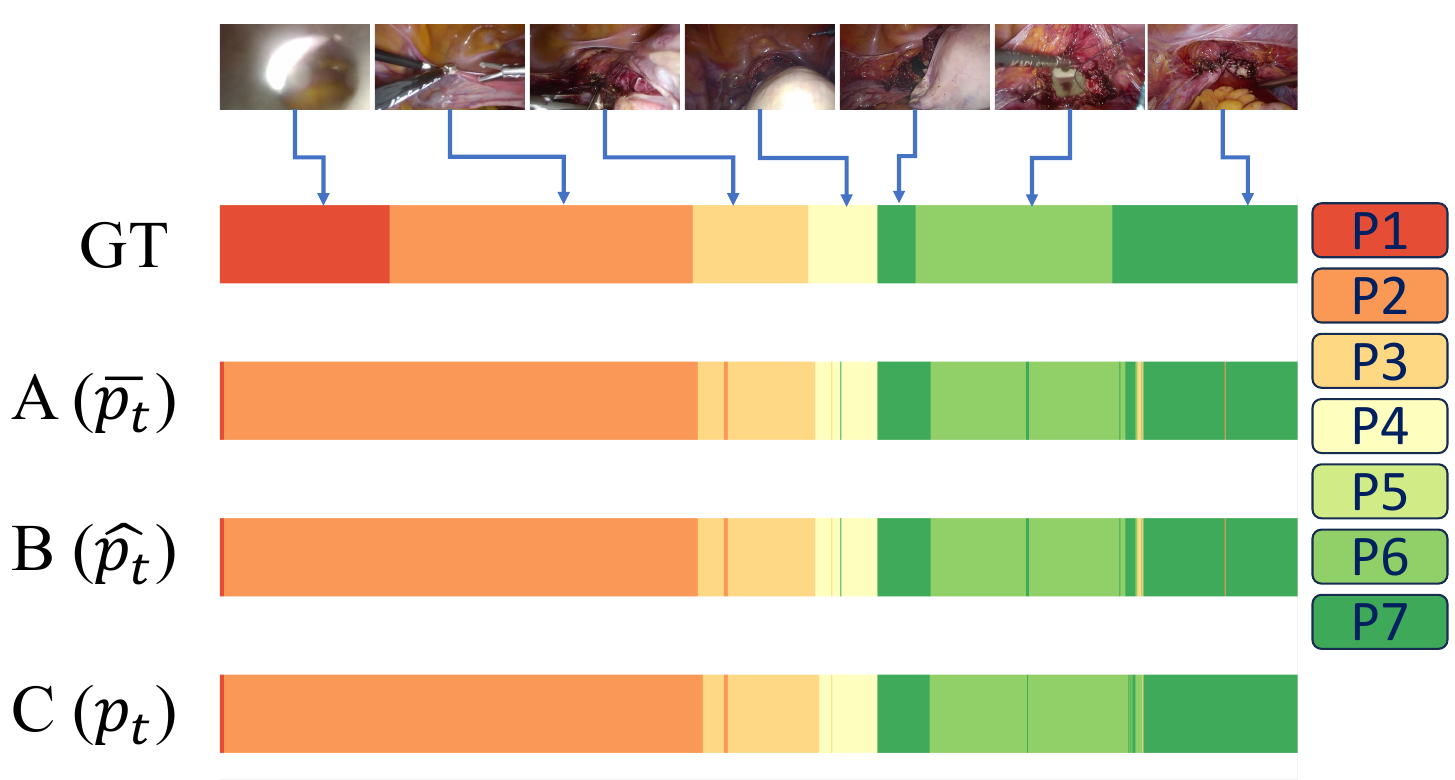}
    \resizebox{0.66\linewidth}{!}{
    \begin{tabular}[b]{c|l|llll}\hline
      \multicolumn{6}{c}{Encoder-decoder on AutoLaparo~\cite{wang2022autolaparo}}\\
      \midrule  
     & {Video-level} & \multicolumn{4}{c}{Phase-level} \\
     & \multicolumn{1}{c|}{Accuracy} & \multicolumn{1}{c}{Precision} & \multicolumn{1}{c}{Recall} & \multicolumn{1}{c}{Jaccard}& \multicolumn{1}{c}{mAP} \\
     \midrule
     A ($\bar{p}_t$) & $81.8$ & $76.5$ & $66.2$ & $56.2$ & 73.1 \\
    B ($\hat{p}_t$) & 82.1 \textsubscript{(+0.3)}
    & 74.2 
    & 67.1 \textsubscript{(+0.9)}
    & 57.2 \textsubscript{(+1.0)}
    & 73.6 \textsubscript{(+0.5)}\\
    C ($p_t$) & 82.9 \textsubscript{(+1.1)}
    & 76.4 
    & 67.9 \textsubscript{(+1.7)}
    & 58.4 \textsubscript{(+2.2)}
    & 74.8 \textsubscript{(+1.7)}\\
    \bottomrule
    \end{tabular}
    }
    \captionlistentry[table]{A table beside a figure}
    \vspace{-1em}
    \caption{Qualitative and quantitative ablation on encoder-decoder architecture. A: baseline prediction $\bar{p}_t$; B: transition-aware prediction $\hat{p}_t$; C: full prediction $p_t$.}
    \label{fig:enc-dec}
  \end{figure}

\noindent\textbf{Implementation Details.}
All our experiments conducted on four NVIDIA V100 GPUs. For frame encoding, we employ Surgformer~\cite{yang2024surgformer} as the backbone and strictly follow their training and testing protocols using their official implementation. For model training, we first train the Surgformer with the original configuration and then add our NSFM module to the original model at the end and keep training for 5 more epochs with a learning rate of $1\times 10^{-5}$.


\subsection{Benchmark Results}
\label{sec:benchmark}
We benchmark against state-of-the-art methods for online surgical phase recognition on all three benchmarks. Results for Surgformer~\cite{yang2024surgformer} are obtained using their official implementation and checkpoints, while other baseline results are from published papers or reproductions in~\cite{liu2023skit} and ~\cite{Lavanchy2024}. On more challenging benchmarks AutoLaparo~\cite{wang2022autolaparo} and BernBypass70~\cite{Lavanchy2024}, NFSM enhances performance across most metrics for both datasets and achieves state-of-the-art performance across most metrics on both datasets, validating both NFSM's generalizability and its capacity to enhance existing architectures. On Cholec80~\cite{TwinandaSMMMP17endonet} benchmark, our NSFM slightly improves the baseline model on metrics other than \textit{mAP}. This is because the dataset is relatively easy so in most of the cases the local information is enough to discriminate the right labels. While NFSM shows noticeable improvement (+1.7) on mAP that considers confidence level, indicating a better alignment of confidence and accuracy.

\subsection{Ablation Study}
\label{sec:components}


In this section, we demonstrate the effectiveness of the proposed Neural Finite-State Machine (NFSM) module by examining three factors: i) \emph{transition-aware training}, ii) \emph{transition-aware inference} (i.e., dynamic transition table utilization), and iii) the adaptability and generalizability of NFSM to other architectures.

\noindent\textbf{Effectiveness.}\quad
We begin with Surgformer as our baseline model, yielding a baseline prediction \(\bar{p}_t\) for the current frame \(t\). We then progressively incorporate (1) \emph{transition-aware training}, producing a transition-aware prediction \(\hat{p}_t\), and (2) \emph{transition-aware inference}, producing the final prediction \(p_t\). We evaluate on two challenging benchmarks, AutoLaparo~\cite{wang2022autolaparo} and BernBypass70~\cite{Lavanchy2024}, using the same metrics from Section~\ref{sec:benchmark}. As shown in Table~\ref{tab:components}, implementing NFSM training yields consistent improvements across most metrics, indicating that the \emph{transition-aware training} approach effectively refines feature representations for more accurate predictions. Incorporating \emph{transition-aware inference} further boosts phase-level precision and \emph{mAP}, underscoring the model’s capacity to leverage both current and historical context for improved future prediction.

\noindent\textbf{Adaptability.}\quad
We further demonstrate NFSM’s adaptability in scenarios where fine-tuning the baseline model may be infeasible—such as when using large foundation models or proprietary architectures that cannot be fully accessed. In our experiments, we freeze all layers in Surgformer and train only the NFSM module. The results in the last two rows of Table~\ref{tab:components} show that even without fine-tuning the original model, the NFSM module alone contributes noticeable performance gains. These findings confirm that NFSM can be employed as a plug-and-play solution to adapt foundation models to specific surgical applications, without requiring extensive re-training or modifications to the underlying architecture.

\noindent\textbf{Generalizability.}\quad
Finally, we validate the NFSM’s generalizability by integrating it into a two-stage encoder-decoder framework, using MViTv2~\cite{LiW0MXMF22mvitv2} as the encoder and RetNet~\cite{SUN23Retnet} as the decoder. Evaluated on the AutoLaparo~\cite{wang2022autolaparo} dataset with the same metrics, this encoder-decoder setup also demonstrates consistent improvements from both NFSM training and inference. Qualitative results in Figure~\ref{fig:enc-dec} illustrate that applying NFSM yields more stable predictions over time, reducing fragmentation within phases and mitigating local perturbations. These outcomes confirm the NFSM’s capacity to generalize across diverse architectures while providing consistent gains in surgical phase recognition.

\section{Conclusion}
\label{section:conclusion}
In this paper, we introduced the Neural Finite-State Machine (NFSM) module, uniting traditional finite-state machine principles with modern deep learning paradigms for surgical phase recognition. By leveraging learnable global state embeddings and dynamic transition tables, NFSM encodes procedure-level knowledge and supports online prediction through transition-aware mechanisms. When integrated into state-of-the-art architectures, NFSM substantially boosts recognition performance by combining both immediate predictions and learned transitions. Our ablation studies confirm the effectiveness of each component, underscoring the module’s generalizability and adaptability across various settings.

Looking ahead, foundation models have the potential to deliver broad, general understanding of surgical video content. Building on this, the NFSM, with strong generalizability and adaptability, can serve as a specialized plug-in module that integrates procedure-specific domain knowledge, further improving performance for particular surgeries. By merging the interpretability of finite-state models with the flexibility of neural networks, NFSM provides a robust pathway for procedure-specific video analysis and sets the stage for integrating structured domain knowledge into future deep learning frameworks.

%
%
%
\bibliographystyle{splncs04}
\bibliography{main}

\begin{thebibliography}{10}
\providecommand{\url}[1]{\texttt{#1}}
\providecommand{\urlprefix}{URL }
\providecommand{\doi}[1]{https://doi.org/#1}

\bibitem{BhatiaOXH07RL_video_hmm}
Bhatia, B., Oates, T., Xiao, Y., Hu, P.F.: Real-time identification of operating room state from video. In: In Proc. AAAI (2007)

\bibitem{BlumFN10}
Blum, T., Feu{\ss}ner, H., Navab, N.: Modeling and segmentation of surgical workflow from laparoscopic video. In: In Proc. MICCAI (2010)

\bibitem{BlumPFN08HMM}
Blum, T., Padoy, N., Feu{\ss}ner, H., Navab, N.: Modeling and online recognition of surgical phases using hidden markov models. In: In Proc. MICCAI (2008)

\bibitem{CadeneRTC16CNN_HMM_SMOOTHING}
Cad{\`{e}}ne, R., Robert, T., Thome, N., Cord, M.: {M2CAI} workflow challenge: Convolutional neural networks with time smoothing and hidden markov model for video frames classification. CoRR  \textbf{abs/1610.05541} (2016)

\bibitem{czempiel2020tecno}
Czempiel, T., Paschali, M., Keicher, M., Simson, W., Feussner, H., Kim, S.T., Navab, N.: Tecno: Surgical phase recognition with multi-stage temporal convolutional networks. In: In Proc. MICCAI (2020)

\bibitem{ding2024towards}
Ding, H., Zhang, Y., Shu, H., Lian, X., Kim, J.W., Krieger, A., Unberath, M.: Towards robust algorithms for surgical phase recognition via digital twin-based scene representation. arXiv preprint arXiv:2410.20026  (2024)

\bibitem{DosovitskiyB0WZ21vit}
Dosovitskiy, A., Beyer, L., Kolesnikov, A., Weissenborn, D., Zhai, X., Unterthiner, T., Dehghani, M., Minderer, M., Heigold, G., Gelly, S., Uszkoreit, J., Houlsby, N.: An image is worth 16x16 words: Transformers for image recognition at scale. In: In Proc. ICLR (2021)

\bibitem{funke2023metrics}
Funke, I., Rivoir, D., Speidel, S.: Metrics matter in surgical phase recognition. arXiv preprint arXiv:2305.13961  (2023)

\bibitem{GaoJDH20treesearch}
Gao, X., Jin, Y., Dou, Q., Heng, P.: Automatic gesture recognition in robot-assisted surgery with reinforcement learning and tree search. In: In Proc. ICRA (2020)

\bibitem{GaoJLDH21transsvnet}
Gao, X., Jin, Y., Long, Y., Dou, Q., Heng, P.: Trans-svnet: Accurate phase recognition from surgical videos via hybrid embedding aggregation transformer. In: In Proc. MICCAI (2021)

\bibitem{GirdharG21avt}
Girdhar, R., Grauman, K.: Anticipative video transformer. In: In Proc. ICCV (2021)

\bibitem{hochreiter1997long}
Hochreiter, S., Schmidhuber, J.: Long short-term memory. Neural computation  \textbf{9}(8),  1735--1780 (1997)

\bibitem{JinD0YQFH18sv-rcnet}
Jin, Y., Dou, Q., Chen, H., Yu, L., Qin, J., Fu, C., Heng, P.: Sv-rcnet: Workflow recognition from surgical videos using recurrent convolutional network. {IEEE} Trans. Medical Imaging  (2018)

\bibitem{JinLDCQFH20mrcnet}
Jin, Y., Li, H., Dou, Q., Chen, H., Qin, J., Fu, C., Heng, P.: Multi-task recurrent convolutional network with correlation loss for surgical video analysis. Medical Image Anal.  (2020)

\bibitem{JinLCZDH21tmrnet}
Jin, Y., Long, Y., Chen, C., Zhao, Z., Dou, Q., Heng, P.: Temporal memory relation network for workflow recognition from surgical video. {IEEE} Trans. Medical Imaging  (2021)

\bibitem{KilleenZMATOU23_Pelphix}
Killeen, B.D., Zhang, H., Mangulabnan, J., Armand, M., Taylor, R.H., Osgood, G., Unberath, M.: Pelphix: Surgical phase recognition from x-ray images in percutaneous pelvic fixation. In: In Proc. MICCAI (2023)

\bibitem{Lavanchy2024}
Lavanchy, J.L., Ramesh, S., Dall’Alba, D., Gonzalez, C., Fiorini, P., M\"{u}ller-Stich, B.P., Nett, P.C., Marescaux, J., Mutter, D., Padoy, N.: Challenges in multi-centric generalization: phase and step recognition in roux-en-y gastric bypass surgery. International Journal of Computer Assisted Radiology and Surgery  (2024)

\bibitem{LeaVRH16tcn}
Lea, C., Vidal, R., Reiter, A., Hager, G.D.: Temporal convolutional networks: {A} unified approach to action segmentation. In: ECCV Workshops (2016)

\bibitem{LiW0MXMF22mvitv2}
Li, Y., Wu, C., Fan, H., Mangalam, K., Xiong, B., Malik, J., Feichtenhofer, C.: Mvitv2: Improved multiscale vision transformers for classification and detection. In: In Proc. CVPR (2022)

\bibitem{Liu23Lovit}
Liu, Y., Boels, M., Garc{\'{\i}}a{-}Peraza{-}Herrera, L.C., Vercauteren, T., Dasgupta, P., Granados, A., Ourselin, S.: Lovit: Long video transformer for surgical phase recognition. CoRR  \textbf{abs/2305.08989} (2023)

\bibitem{liu2023skit}
Liu, Y., Huo, J., Peng, J., Sparks, R., Dasgupta, P., Granados, A., Ourselin, S.: Skit: a fast key information video transformer for online surgical phase recognition. In: Proc. ICCV (2023)

\bibitem{PadoyBAFBN12}
Padoy, N., Blum, T., Ahmadi, S., Feu{\ss}ner, H., Berger, M., Navab, N.: Statistical modeling and recognition of surgical workflow. Medical Image Anal.  (2012)

\bibitem{PadoyBFBN08_online}
Padoy, N., Blum, T., Feu{\ss}ner, H., Berger, M., Navab, N.: On-line recognition of surgical activity for monitoring in the operating room. In: In Proc. AAAI (2008)

\bibitem{SUN23Retnet}
Sun, Y., Dong, L., Huang, S., Ma, S., Xia, Y., Xue, J., Wang, J., Wei, F.: Retentive network: {A} successor to transformer for large language models. CoRR  \textbf{abs/2307.08621} (2023)

\bibitem{TwinandaSMMMP17endonet}
Twinanda, A.P., Shehata, S., Mutter, D., Marescaux, J., de~Mathelin, M., Padoy, N.: Endonet: {A} deep architecture for recognition tasks on laparoscopic videos. {IEEE} Trans. Medical Imaging  (2017)

\bibitem{valderrama2020tapir}
Valderrama, N., Puentes, P.R., Hern{\'a}ndez, I., Verlyk, N.A.M., Santander, J., Caicedo, J., Fern{\'a}ndez, N., Arbel{\'a}es, P.: Towards holistic surgical scene understanding. In: In Proc. MICCAI (2022)

\bibitem{VaswaniSPUJGKP17transfomer}
Vaswani, A., Shazeer, N., Parmar, N., Uszkoreit, J., Jones, L., Gomez, A.N., Kaiser, L., Polosukhin, I.: Attention is all you need. In: In Proc. NIPS (2017)

\bibitem{wang2022autolaparo}
Wang, Z., Lu, B., Long, Y., Zhong, F., Cheung, T.H., Dou, Q., Liu, Y.: Autolaparo: A new dataset of integrated multi-tasks for image-guided surgical automation in laparoscopic hysterectomy. In: In Proc. MICCAI (2022)

\bibitem{yang2024surgformer}
Yang, S., Luo, L., Wang, Q., Chen, H.: Surgformer: Surgical transformer with hierarchical temporal attention for surgical phase recognition. In: In Proc. MICCAI (2024)

\bibitem{YiJ19hfom}
Yi, F., Jiang, T.: Hard frame detection and online mapping for surgical phase recognition. In: In Proc. MICCAI (2019)

\bibitem{zhao2022real}
Zhao, Y., Kr{\"a}henb{\"u}hl, P.: Real-time online video detection with temporal smoothing transformers. In: Proc, ECCV (2022)

\end{thebibliography}

\end{document}